\documentclass[a4paper,fleqn,runningheads]{llncs}

\usepackage[T1]{fontenc}
\usepackage[utf8]{inputenc}
\usepackage{lmodern}
\usepackage{microtype}

\usepackage{pifont}
\usepackage{ellipsis}
\usepackage{cite}
\usepackage{verbatim}
\usepackage{fancyvrb}
\usepackage{listings}
\usepackage{amssymb}
\usepackage[fleqn]{amsmath}
\usepackage{amsfonts}
\usepackage{amsbsy}
\usepackage[fleqn,amsmath]{ntheorem}
\usepackage{dsfont}
\usepackage{latexsym}
\usepackage{xcolor}
\definecolor{linkcolor}{RGB}{31,31,222}

\usepackage[ruled,linesnumbered,noline]{algorithm2e}
\usepackage[noend]{algpseudocode}

\usepackage[pdftex,colorlinks,linkcolor=linkcolor,citecolor=linkcolor,urlcolor=linkcolor,linktocpage]{hyperref}
\usepackage[nameinlink]{cleveref}
\usepackage{makeidx}
\usepackage[format=plain,labelsep=period,font=footnotesize,labelfont=bf,skip=2mm]{caption}[2004/07/16]
\usepackage{subfig}
\usepackage{courier}
\usepackage{amstext}
\usepackage{relsize}

\usepackage{enumerate}
\usepackage{multirow}
\usepackage{paralist}
\usepackage{enumitem}

\usepackage{booktabs}
\usepackage{array}
\usepackage{xspace}

\usepackage{graphicx}
\usepackage{tikz}
\usepackage{pgfplots}

\let\leftold\left
\let\rightold\right
\renewcommand{\left}{\mathopen{}\mathclose\bgroup\leftold}
\renewcommand{\right}{\aftergroup\egroup\rightold}

\crefname{definition}{Definition}{Definitions}
\crefname{theorem}{Theorem}{Theorems}
\crefname{lemma}{Lemma}{Lemmata}
\crefname{observation}{Observation}{Observations}
\crefname{conjecture}{Conjecture}{Conjectures}
\crefname{proposition}{Proposition}{Propositions}
\crefname{figure}{Figure}{Figures}
\crefname{table}{Table}{Tables}
\crefname{section}{Section}{Sections}
\crefname{subsection}{Subsection}{Subsections}
\crefname{subsubsection}{Subsection}{Subsections}
\crefname{algorithm}{Algorithm}{Algorithms}
\crefname{note}{Note}{Notes}
\crefname{enumi}{}{}
\crefname{equation}{}{}

\theoremseparator{.}

\begin{document}

\newcommand{\cancel}[1]{}

\title{Second Thoughts on the Second Law}

\author{Stefan Wolf}

\institute{Faculty of Informatics, Universit\`a della Svizzera
  italiana, 6900 Lugano, Switzerland\\
{\it Facolt\`a indipendente di Gandria, 6978 Gandria, Switzerland}}

\maketitle

\begin{abstract}
We speculate whether  the {\em second law of thermodynamics\/}
has more to do with Turing machines than steam pipes. It
states the {\em logical reversibility\/} of reality as a
computation, {\em i.e.}, the fact that no information is 
forgotten:
nature computes with Toffoli-, not NAND gates. 
 On the way there,
we correct Landauer's erasure principle
by directly linking it 
to  {\em lossless data compression}, and  we further develop that to a
lower bound on the 
energy
 consumption and heat dissipation  of  a general computation.
\end{abstract}

\section{Prologue}

A few years ago, our group  had the great pleasure and privilege to receive 
{\em Juraj Hromkovi\v{c}\/} as a guest 
in {\em Ticino}. He had announced a discourse on the topic {\em ``What is information?''}. It was a
fascinating lecture in which Juraj was advocating to view 
information
as {\em complexity}. This  has been inspiring for me, and it is probably 
not a co\"{\i}ncidence that soon after that, I realized that the use
of {\em Kolmogorov complexity\/}~\cite{text}\footnote{The Kolmogorov
  complexity $K_{\cal U}(x)$ of a string $x$ with respect to a
  universal Turing machine ${\cal U}$ is the length of the shortest
  program for ${\cal U}$ that  outputs $x$. 
The {\em conditional\/} complexity of $x$ given $y$, $K_{\cal U}(x|y)$, is the
length of the shortest program outputting~$x$ upon input $y$.
The quantities depend
  on the choice of the specific machine  ${\cal U}$ only through an additive constant.} 
instead of probability distributions in 
the context of the fascinating but strange ``non-local'' correlations that quantum physics 
comes with~--- and that had been my main object of study for over a
decade already at that time~--- offers a significant conceptual
advantage: {\em Non-counterfactuality}, {\em i.e.}, no 
 necessity 
to talk about the
outcomes 
of unperformed experiments. 

{\em John Stewart
Bell\/} showed in 1964~\cite{bell} that quantum theory predicts
correlations between measurement outcomes that are too strong to be
explained by shared classical information. This is as if identical
twins did not only {\em look\/} alike~--- such  correlations can easily be
explained with their identical DNA sequences and do not confront 
us with a ``metaphysical'' problem~--- but also {\em behaved\/} in a
ways so strongly correlated that
genetic explanations fail. Bell's result
was a late reply to an attack to quantum theory by {\em Einstein, Podolsky, and Rosen\/}
(``EPR'') in 1935~\cite{EPR}
who remarked that
if the outcomes of measurements on quantum systems are
correlated, 
then they cannot be spontaneously random as
predicted by the theory,
but they must already have been determined beforehand, at the occasion
of the creation
of the {\em entangled pair}. To stay with our analogy: If twins
look the same, their looks must  be determined by their
genes; if they also behave the same, then so must their behaviour. This is a natural thought~--- but it is insufficient, and to
realize this is Bell's breakthrough: {\em ``Genetic'' explanations are too weak 
for quantum correlations.} But then, where {\em do\/} they  come from? 
What mechanism establishes them?

The basis of {\em EPR\/}'s argument
has later been called 
 ``Reichenbach's
principle''~\cite{Reichenbach}:
A~correlation in a causal structure is established either by a
{\em common cause\/} in the common past or a {\em direct influence\/}
from one event to the other.  Bell's celebrated argument rules out the first possibility 
if that
common cause is to be a piece of classical information. 
{\em Influence\/} stories for establishing quantum correlations
cannot be ruled out entirely, but they 
require the speed of that influence to be infinite, and they are
unnatural in other respects expressing the fact that  explaining a {\em non\/}-signaling
phenomenon with a {\em signaling\/} mechanism is 
shooting sparrows with cannons.
Eventually, {\em the fundamentality of the 
causal structure is in question\/}\footnote{The idea of dropping
causality may sound radical but  is not new; see {\em Bertrand
  Russell, 1913\/}~\cite{russell}: ``The law of causality [\ldots]\ is a relic of a bygone age, surviving,
like the monarchy, only because it is erroneously supposed to do no
harm.''}~--- the only assumption for   Reichenbach's
principle. So motivated, models of {\em relaxed causality\/} have  been studied~\cite{ocb}
and disclosed an unexpectedly rich world~\cite{nparties} between fixed causal orders
and logical inconsistency ({\em \`a la\/} ``grandfather paradox''~--- you
travel to the past and kill your own grandfather~--- {\em etc.\/}),
much like {\em Bell world\/} between locality and signaling. 

The fall of rigid causality comes with
a further victim, {\em randomness\/}: Common physical
definitions of freeness of randomness~\cite{colrenamp}
are based on that very structure and  they fall with it. One way out
is to consider freeness of randomness as fundamental and
causality as
emerging from it, via: {\em ``What is correlated with a perfectly free
  bit must be
in its future.''\/} For single bits, this is the best we can hope for.
For bit {\em strings}, however, there can be an {\em intrinsic
  randomness notion\/} depending only on the data itself but not (otherwise) on 
the process leading up to them. In~our search
for such a non-contextual view,
 we land in  a field
traditionally  related to probability distributions and
ensembles
but that knows a ``non-counterfactual'' viewpoint as well: {\em
  Thermodynamics}.

\section{From the Steam Pipe\/ \ldots}

The story\footnote{Most of the following historical facts are drawn from the article 
``Bluff your way in the second law of thermodynamics'' by {\em Jos Uffink},
2001.}
of the {\em second law of thermodynamics\/}
 starts with {\em Sadi Carnot\/}
(1796--1832) and
his
study of heat engines such as {\em James Watt\/}'s steam pipe. 
To conclude that the law
manifests itself only for
such 
engines and  their circular processes 
means
 to underestimate
a   general {\em combinatorial fact}.
The second law was discovered through steam
engines because they first  permitted a precise unclouded view on it~--- but
 the law is restricted to combustion engines  as little as Jupiter's moons
 depend on telescopes.\footnote{A symptom of the
  law's generality is that it has advanced to
becoming pop
culture in the meantime: see, {\em e.g.}, Allen, W., {\em Husbands and Wives\/}
  (1992)~--- The protagonist
  Sally is explaining why her marriage did not work out. First she
  does not know, then she realizes: ``It's the second law of
  thermodynamics: sooner or later, everything turns to shit. That's my
  phrasing, not the {\em Encyclopedia Britannica}.'' 
The second law is  less popular than Einstein's 
elegant
relativity or 
Bennett {\em et al.\/}'s  sexy
 teleportation
since it does not have 
any glamour, fascination,  or promise  attached to it:
Quite on the contrary, it stands for
facts we  usually    deny or try to avoid.}

{\em Carnot\/} discovered
 that the maximal efficiency of  a heat engine between 
two heat baths depended solely on the two temperatures involved.
This was his only publication;
it appeared when he was 28 and was entitled: {\em
  ``R\'eflexions sur la 
puissance motrice du feu et sur les machines propres à d\'evelopper cette puissance.''}

{\em Rudolf Clausius'\/} (1822--1888) version of the second law reads: {\em ``Es kann nie W\"arme aus einem k\"alteren in einen
w\"armeren 
K\"orper \"ubergehen, ohne dass eine andere damit zusammenh\"angende 
\"Anderung eintritt.''}~--- ``No process can transport
heat 
from cold to hot and do no further change.''

{\em W.\ Thomson (Lord Kelvin)\/} (1824--1907)
formulated his own version of the second law and then concluded that the
law may have consequences more severe than what is obvious at first sight:
{\em ``Restoration of mechanical energy without dissipation
[...]\ is impossible.
Within a finite period of time past, the earth must have been, within
a finite time, the earth must again be unfit for the habitation of man.''}

Also for Clausius, it was only a single thinking step
from his version of the law to conclude that 
all temperature differences in the entire universe will vanish~--- the
{\em ``W\"armetod''\/}~--- and
that then, no
change will happen anymore. 
He speaks of  a general tendency of nature for change into
  a specific direction: {\em ``Wendet man
dieses
auf das Weltall im Ganzen an, so gelangt man zu einer eigent\"umlichen
Schlussfolgerung, auf welche zuerst W.~Thomson aufmerksam
machte,
 nachdem er sich meiner Auffassung des zweiten Hauptsatzes
angeschlossen hatte.\footnote{Roughly: ``He had an interesting view on
  the second law
  after having adopted mine.''}
Wenn [...]\ im Weltall die W\"arme stets das
Bestreben zeigt, [...]\ dass [...]\ Temperaturdifferenzen
ausgeglichen werden, so muss es sich mehr und mehr dem Zustand
ann\"ahern, wo [...]\ keine Temperaturdifferenzen mehr 
existieren.''\/}~--- ``When this is applied to the universe as a whole,
one gets to the strange conclusion that already W.~Thomson had pointed
out after having  taken my view of the second law. If
heat always tends towards reducing temperature differences, then the universe will
approximate more and more  the state in which no temperature
differences exist anymore.''\footnote{It seems that his faithful pupil {\em Max Planck\/}
believed that claim was untenable~--- Clausius finally erased, last-minute, all remarks 
concerning  ``the entropy of the universe as a whole'' from his collected
works~--- by hand.}

{\em Ludwig Boltzmann\/} (1844--1906) brought our understanding of the second 
law closer to combinatorics and probability theory:
 The second law was for him the expression of the fact 
that it is more likely to end up in a {\em large\/} set of possible states
than in a small one:
The higher the number of 
{\em particular\/} situations (microstates) which belong to a 
{\em general\/} one (macrostate), the more likely it is
to be in that general situation. In other 
words,
  time evolution does not decrease a closed system's {\em entropy},
  which is
proportional to the logarithm of the number of
corresponding microstates: Things do not get ``more special'' with time.\footnote{The
  corresponding
  formula $S=k\ln W$ is written in golden letters on Boltzmann's gravestone at the 
{\em Zentralfriedhof\/}~--- which is, as the word goes in Vienna,
``halb so gross und doppelt so lustig wie Z\"urich''.
Boltzmann  imagined
that the universe had started in a completely ``uniform'' state.
So the diverse reality now perceived would  be a mere
fluctuation. Note that the fact that this fluctuation is extremely
  unlikely is irrelevant if you can {\em condition on our existence},
    given your thinking all this (this thought is sometimes called ``the anthropic principle'').
He was probably  aware that this way of thinking may lead 
into {\em solipsism:\/}  ``My existence alone, simply imagining all that, is much more likely than the actual
existence of all people around me, let alone all the visible galaxies,
{\em etc}.'' He eventually hung himself in a hotel room in Duino,
Italy. It has
 been colported in Vienna that this may have been due to  ``mobbing'' at the
university by Ernst Mach. Anyhow, today we prefer to comfort ourselves 
with the contrary belief that the universe initiated in a low-entropy
state,
and we call this assumption
 ``the big bang.''} 

Boltzmann's notions of  {\em macrostate\/} and {\em entropy\/} are subjective, and it is not  
obvious how to define them in general {\em e.g.}, for
non-equilibria. We 
propose instead a version of the second law 
that is 
broader and  more precise  at the same time,  
 avoiding probabilities and ensembles. Crucial steps in that direction were made by {\em Wojciech Zurek}~\cite{zurek}. 
We follow  {\em Rolf Landauer\/}~\cite{landauer61} whose choice of  viewpoint about 
thermodynamics can be compared with {\em Ernst Specker\/}'s~\cite{specker} about 
quantum theory: {\em Logic.}

\section{\ldots to the Turing Machine}

Landauer~\cite{landauer61} investigated the thermodynamic price of logical 
operations. He was correcting a belief by {\em John von Neumann\/}
that 
every bit operation required free energy $kT\ln 2$:\footnote{Here, $k$ is
Boltzmann's 
constant connecting the micro- and macroscopic realms, $T$ the environmental temperature~--- and the factor $\ln2$
is a common sight at the border between logic and physics with their
respective basic constants $2$ and $e$.} 
According to Landauer~--- as confirmed by {\em Fredkin and Toffoli's\/}~\cite{ballist}
``ballistic computer''~---,
that
price is unavoidable only for 
 {\em logically irreversible\/}
operations such as the AND or the OR. On
the positive side, it has been observed~\cite{bennetttoc} that
every function, bijective or not, can in principle be evaluated in a
{\em logically  reversible way,
  using
only ``Toffoli gates,'' i.e., made-reversible and then-universal AND gates\/};  this
computation can then be thermodynamically neutral, {\em i.e.}, it
 does not  dissipate heat.

{\em Landauer's principle\/} states that the erasure (setting the
corresponding binary memory cells to $0$) of $N$ bits costs 
$kTN\ln2$ free energy which must be dissipated as heat to the 
environment, a thermal bath of temperature $T$. The {\em dissipation\/} is crucial in the argument:
Heating up the environment compensates
for the {\em loss of entropy\/} within the memory cell which is
materialized by some physical system (spin, gas molecule, {\em etc.}). Landauer's 
principle is a direct consequence of Boltzmann's view of the second
law.

{\em Charles Bennett\/}~\cite{bennettLP} used Landauer's slogan ``Information is
Physical'' for making the key contribution to
the resolution of the paradox of ``Maxwell's
demon''~(see, {\em e.g.},~\cite{szilard29}). 
That demon 
had been thought of as violating the second law\footnote{Also other
  authors noted the strange dependency of the law on the absence of
  certain life forms~--- Kelvin wrote: ``When light is absorbed {\em other 
than in vegetation}, there is dissipation [...]'').
To make things worse, there is always a non-zero probability  (exponentially
small, though) of
exceptions,
where the law fails to hold; we  are not used to this from other
laws of physics (outside quantum theory). 
This {\em fragility\/} of the second law is weirdly
contrasted by
its being, at the same time, {\em more robust\/} than others, such as Bell 
violations only realizable in extreme lab conditions: We  do not need 
to trust experimentalists to be convinced
that ``it'' is there --- in fact everywhere.}
by adaptively 
handling a frictionless door with the goal of ``sorting a gas'' in
a container. Bennett took 
 the demon's memory (imagined to be in the all-$0$-state before sorting)
into account, which is in the end filled with
``random'' information remembering the original state of the
gas: The 
growth of disorder {\em within\/} the demon compensates for the order 
she creates {\em outside} ({\em i.e.}, in the gas)~--- the second  law is
saved. The initial $0$-string is the demon's resource
allowing for her order creation (see Figure~\ref{fig:2}):
\begin{figure}[h]
	\centering
	\includegraphics[scale=0.3]{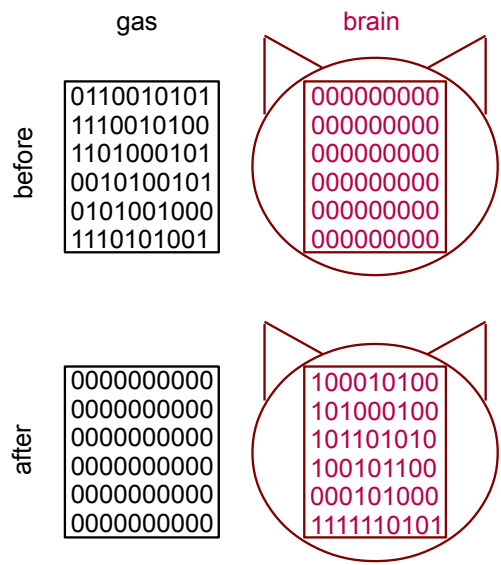}
	\caption{Bennett's resolution of the Maxwell-demon paradox.}
	\label{fig:2}
\end{figure}
If we break Bennett's argument apart in the middle, we  end up with
the {\em converse\/} of Landauer's principle: 
 The 
all-$0$-string has {\em work value}, {\em i.e.}, if we accept 
 the respective memory cells to become ``randomized,'' 
we can  extract $kTN\ln2$ free energy from 
the environment (of temperature $T$).

Already Bennett~\cite{bennetttoc} had implied that for some strings $S$,
the {\em erasure cost\/}  is less than Landauer's 
len$(S)\cdot kT\ln 2$: Besides the obvious
$00\cdots 0$ and $11\cdots 1$, this is also true, {\em e.g.}, for the
string formed by the first $N$ digits of the binary expansion of
$\pi$: The reason is that there is a {\em short program\/} generating the 
sequence or, in other words, a logically {\em reversible\/}
computation between (essentially) $0^N$  and $\pi^N$ that can be carried out 
thermodynamically reversibly~\cite{ballist}. Generally, if the string can be
{\em compressed\/} in a lossless fashion, then  the erasure cost shrinks accordingly. 

\begin{figure}[h]
	\centering
	\includegraphics[scale=0.3]{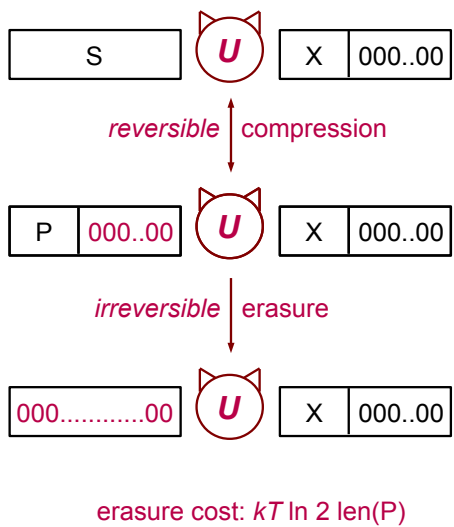}
	\caption{Erasing $S$ with catalyst $X$ at cost EC$_{\cal
            U}(S|X)$: First, $X$ is reversibly compressed to the
          shorter string $P$ for free, then that ``program'' $P$ is
          erased at cost $kT\ln 2\cdot {\rm len}(P)$.}
	\label{fig:1}
\end{figure}

Let us consider a {\em model for the erasure process\/} (see Figure~\ref{fig:1})
in which there is, besides the string~$S$ to be erased, another string
$X$ on the tape as a ``catalyst'' 
summarizing possible {\em a~priori\/} ``knowledge'' about~$S$,
 so helping to compress it
but remaining itself unchanged in the process. The universal Turing
machine~${\cal U}$'s tape is assumed to be finite~--- there would be 
infinite ``work value'' on it otherwise~---, and the resulting erasure cost
is~EC$_{\cal U}(S|X)$.

Bennett~\cite{bennetttoc} claims that the erasure cost of a string is,
actually, its Kolmogorov complexity times $kT\ln 2$; this would in our model translate
to the erasure cost equalling the conditional complexity
of the string to be erased, 
 given the side information: EC$_{\cal U}(S|X)=K_{\cal U}(S|X)$. 
Unfortunately, this is in general not achievable due to the {\em
  uncomputability\/} of Kolmogorov complexity and the corresponding
compression transformation, and since we 
assume the extraction to be carried out by a Turing machine, in the spirit of the
{\em Church/Turing hypothesis\/}~\cite{CT}. It is, however,  true
that complexity leads to a {\em lower bound\/} on the erasure cost since 
it represents the ultimate limit on the lossless compressibility of the
string by ${\cal U}$. In the same way, we can argue that any {\em concrete\/} and
{\em computable\/} compression algorithm (with side information) $C$, {\em
  e.g.}, Lempel/Ziv~\cite{zl}, leads
to an {\em upper bound\/} on the erasure cost: First, we reversibly compress
(at no energy cost) and then erase the compression.\footnote{Our result is in a certain contrast
  with claims by Dahlsten {\em et al.\/}~\cite{dahlsten} who take an
  {\em entropic\/} stand: If a ``demon'' knows perfectly the string to
  be erased, this can be done for free. In our algorithmic view, this 
is not reproduced for the case where that knowledge is {\em
  non-constructive}, {\em e.g.}, ``$S$ consists of the first $N$ bits $\Omega^N$ of
the {\em halting probability\/}  $\Omega$ of ${\cal U}$.'' Given that
particular knowledge, the entropy of
$S$ is zero, but still it does not help for erasing $S$ since no
algorithm 
can generate or ``uncompute'' $S$
with the help of that knowledge. In contrast, as long as the 
knowledge is a copy of $S$ or  a program that allows for computing $S$,  the
results of~\cite{dahlsten} and our own match.}

\

\noindent
{\bf Landauer's principle, revisited.}
{\it 
Let $C$ be a computable  function,
\[
C\, :\, \{0,1\}^*\times \{0,1\}^* \rightarrow \{0,1\}^*\ ,
\]
such that 
\[
(V,W)\mapsto (C(V,W),W)
\]
is injective. 
Then the cost of the erasure of $S$ with catalyst $X$,
carried out by  the universal Turing
machine ${\cal U}$, is bounded by
\[
K_{{\cal U}}(S|X)\cdot kT\ln2\ \leq\ {\rm EC}_{\cal U} (S|X)\ \leq\  {\rm len}(C(S,X))
\cdot kT\ln2\ .
\]}

The principle can  be extended to an {\em arbitrary\/} 
computation starting with  input~$A$ and leading up to output $B$ 
with side information $X$,
where $(A,X)$ and $(B,X)$ are the only content of the tape, besides
$0$s, before and after the computation, respectively.
Our result is an algorithmically constructive modification of 
entropic results~\cite{dupuis} and a generalization of less
constructive but also complexity-based claims~\cite{zurekcost}.

\

\noindent
{\bf Landauer's principle, generalized.}
{\it 
Let $C$ be a computable  function,
\[
C\, :\, \{0,1\}^*\times \{0,1\}^* \rightarrow \{0,1\}^*\ ,
\]
such that 
\[
(V,W)\mapsto (C(V,W),W)
\]
is injective. 
Assume that the Turing machine ${\cal U}$ carries out a computation such
that~$A$ and~$B$ are its  initial and final states.
Then the energy cost  of this computation with side information~$X$,
${\rm Cost}_{\cal U}(A\rightarrow B\, |\, X)$, is at least
\[
{\rm Cost}_{\cal U} (A\rightarrow B\, |\, X)\ \geq\  [K_{\cal U}(A|X)-{\rm len}(C(B,X)) ]
\cdot kT\ln2\ .
\]
}\noindent
{\it Proof.}
The erasure cost of $A$, given $X$, is at least $K_{\cal U}(A|X)\cdot kT\ln2$
according to the above. {\em One\/} possibility of realizing this complete
erasure of $A$ is to first transform it to $B$ (given $X$), and then
erase $B$~--- at cost at most len$(C(B,X))\cdot kT\ln2$. Therefore, the
cost to get from $A$ to $B$ given $X$ cannot be lower than the
difference between $K_{\cal U}(A|X)\cdot kT\ln2$ and len$(C(B,X))\cdot
kT\ln2$.
{\em qed.\/}

\

The complexity reductions in these statements quantify the ``amount of logical
irreversibility''
inherent in the respective process, and the quantity of required work~---
  the price for 
 this {\em non-injectivity\/}~---
 is
proportional to that.
The picture is now strangely {\em hybrid\/}: The environment must pay a {\em thermodynamic (macroscopic)\/} 
price for what happens {\em logically (microscopically)\/} in one of
its parts. What prevents us from looking at the environment with a microscope?
If we let ourselves inspire by {\em John Archibald Wheeler\/}'s~\cite{wheeler} ``It
from Bit,'' 
we see the possibility of a compact argument:
The price that the environment has to pay  
in compensation for the irreversibility of the computation in one of 
its parts is such that the {\em   overall computation
is  reversible}.

\

\

\noindent
\fbox{\parbox{12cm}{
\begin{center}
{\bf Second law of Thermodynamics, logico-computational.}\\
{\it 
Time evolutions are logically reversible: No information gets erased.
}\end{center}}}\

\

\

Let us first note that this condition on time evolutions is, like
traditional ``second laws,'' asymmetric in
time: 
Logical reversibility only requires the future to uniquely determine the past, 
 {\em not vice versa\/}: So if 
``reality'' is such an injective
computation, its reverse can be injective as well ({\em e.g.}, computation of a {\em deterministic\/} Turing
machine,  Everett/Bohm interpretation of quantum theory) or this fails
to hold
since the forward direction has
 splitting paths ({\em e.g.}, computation of a {\em probabilistic\/} Turing
machine).

The second law has often been linked to the
  emergence of an {\em arrow of time}. How is that compatible with
  our view?
In  {\em determinism}, 
logical reversibility holds {\em both ways}. What could then  be possible origins of our 
ability to distinguish past and future? (Is it  the limited precision
of our sense organs and the resulting {\em coarse-graining\/}?)
 {\em Indeterminism\/} is easier in that sense  since
it comes with objective asymmetry in time: {\em Randomness\/} points to the future.

\section{Consequences}

\noindent
{\bf Logical reversibility implies quasi-monotonicity.}
\\ \

\noindent
The logical reversibility of a computation implies that the overall
complexity on the Turing machine's tape at time $t$ can be smaller than the one at time~$0$ 
by {\em at most\/} $K(C_t)+O(1)$ if $C_t$ is a string encoding 
the time span $t$. The reason is that one possibility of describing 
the state at time~$0$ is to give the state at time~$t$, plus~$t$
itself; the rest is exhaustive search using only a constant-length
program
simulating forward time evolution. 
\\ \

\

\noindent
{\bf Logical reversibility implies a Boltzmann-like second
   law.}
\\ \

\noindent
The notion of a  {\em macrostate\/} can be defined in an objective way, using 
a ``structure'' function based on the  {\em Kolmogorov sufficient
  statistics\/}~\cite{aare},\,\cite{ccc}.
Roughly speaking, the macrostate is the structure or  ``compressible part'' 
of a microstate: Given the macrostate~--- 
being a set of microstates that is, ideally, small, and its
description at the same time short (such as: ``a gas of volume $V$, temperature~$T$, and 
pressure~$p$ in equilibrium'')~---, the particular microstate is a 
``typical element'' of it, specifiable only through stubborn binary coding. 
This notion of a macrostate is typically unrelated to the
second law except  when the initial and final macrostates 
both have very short descriptions, like for equilibria:  Then, logical reversibility implies that their
size is essentially non-decreasing in time. This is Boltzmann's law. 
\\ \

\

\noindent
{\bf Logical reversibility implies a Clausius-like second law.}
\\ \

\begin{figure}[h]
	\centering
	\includegraphics[scale=0.33]{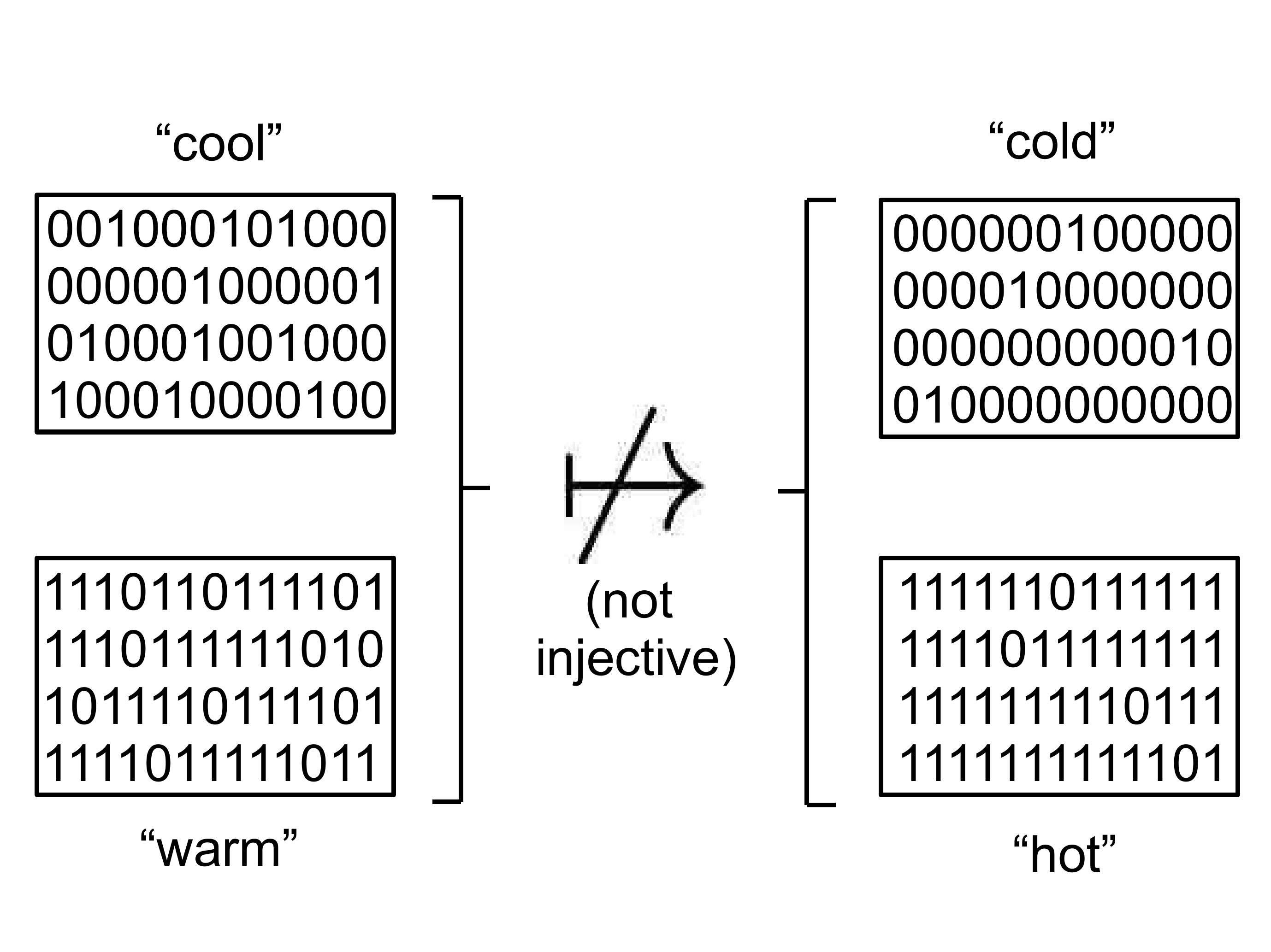}
	\caption{Logical reversibility does not accentuate
          differences: Clausius.}
	\label{claus}
\end{figure}

\noindent
If we have a circuit ~--- the time
evolution~--- using only logically
reversible Toffoli gates, then it is
{\em impossible\/} that this circuit computes  a transformation 
of the following nature: Any pair of strings~--- one with higher
Hamming weight than
the other~--- is mapped to a pair of equally long strings
where the heavy string has become  heavier and the light one 
lighter.
Such a mapping, which  {\em accentuates\/} differences,
cannot  be injective for counting reasons.
We illustrate this with a toy example (see Figure~\ref{claus}). 
\\ \

\noindent
{\it Example.}
Let a circuit consisting of Toffoli gates map an $N(=2n)$-bit string to 
another string~---  which must then be  $N$-bits long as well, due to 
reversibility. We consider the map separately on the first and second 
halves of the full string. We assume the computed function to be 
conservative, {\em i.e.}, to leave the Hamming weight of the full
string
unchanged.
We  look at the excess of $1$'s in one of the halves (which 
equals the deficit of $1$'s in the other). We observe that the
probability
(with respect to the uniform distribution over all strings of some 
Hamming-weight couple $(wn,(1-w)n)$) of the {\em imbalance
substantially growing\/} is exponentially weak. The key ingredient in the argument 
is the function's  injectivity.
Explicitly, the probability that the weight couple goes from
$(wn,(1-w)n)$ 
to $((w+\Delta)n,(1-w-\Delta)n)$~--- or more extremely~---, for $1/2\leq w<1$ and $0<\Delta\leq
1-w$, is
\[
\frac{{n \choose (w+\Delta)n}{n \choose (1-w-\Delta)n}}
{{n \choose wn}{n \choose (1-w)n}}
=2^{-\Theta(n)}\ :
\]
Logical reversibility is incompatible with the tendency of {\em
  polarization\/} of differences.

\vspace*{-0.5cm}

\noindent
{\bf Logical reversibility implies a Kelvin-like second law.}
\\ \

\noindent
Finally, logical reversibility also implies statements resembling
Kelvin's version of the second law: 
``A single heat bath alone has no work value.''
This, again, follows from a  counting argument:
There exists no reversible circuit that
concentrates redundancy to some pre-chosen
bit positions.

\

\noindent
{\it Example.}
The probability that a fixed circuit maps a string of
length~$N$
 and Hamming weight~$w$ to another such that the first $n$
positions 
contain only~$1$'s, and such that the Hamming weight of the remaining 
$N-n$ positions is $w-n$,
is 
\[
\frac{
{N-n \choose w-n}}{{N \choose w}}=2^{-\Theta(n)}\ .
\]
In a sense, Kelvin's law is a special case of Clausius'
formulation.

\section{Epilogue}

\begin{figure}[h]
	\centering
	\includegraphics[scale=0.35]{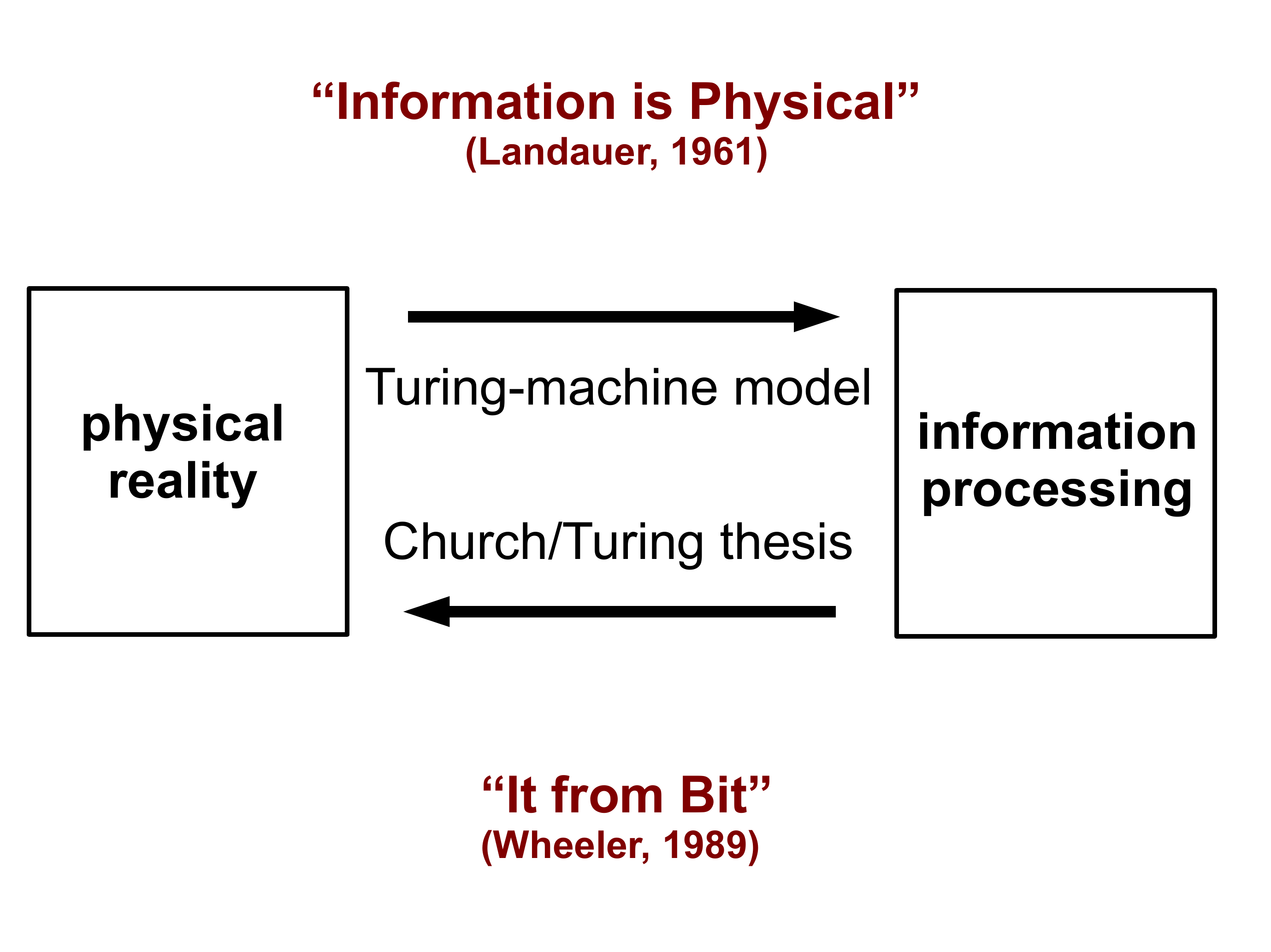}
	\caption{Physics and information: Scenes from a marriage.}
	\label{fig:3}
\end{figure}

{\em A priori}, the relation between physical reality and computation
is two-fold: Turing's famous machine model is  physical~--- 
besides the tape's infiniteness~---, and the
generalized Church/Turing hypothesis suspects  physical processes to
be simulatable on a Turing machine (see Figure~\ref{fig:3}). This
 {\em circulus\/} has
been the background of  our considerations.

\noindent
In the ``Church/Turing view,'' 
a physical law is 
a property of a Turing machine's computation: 
The {\em second law of
    thermodynamics\/} is 
{\em
  logical reversibility}.\footnote{The
  quantum-physical-interpretational reading thereof is the 
Everettian {\em relative-state\/} view: In contrast to collapse
interpretations, it {\em is\/} logically reversible due to its
unitarity.~---
There is a vivid dispute  on such readings of quantum theory that is sometimes more fiery than exact,
more passionate than argumentational: The nature of the debate is ignoring  {\em Paul
Feyerabend\/}'s remark
that in science just as well as in art, a particular {\em style\/} can merely
be judged from the point of view of another, never objectively. 
This call to  modesty culminated 
in seeing {\em science  as  one style among many\/}: ``Man
entscheidet sich f\"ur oder gegen die {\em Wissenschaften\/} genau so, wie
man sich f\"ur oder gegen {\em
  punk rock\/} entscheidet''~\cite{fey}.}

What can be said about the validity of the Church/Turing hypothesis? 
Nothing absolute, just as on
``determinism.'' 
But exactly like for that latter problem, an 
{\em all-or-nothing\/}
statement
creates, at least, a clear dichotomy~\cite{tamc},\,
\cite{charles}.

\

\

\noindent
\fbox{\parbox{12cm}{
\begin{center}
{\bf All-or-Nothing Feature of the Church/Turing Thesis.}\\
{\it 
Either
no entity  can generate non-Turing-computable sequences,\\or 
 even single photons can.
}\end{center}}}\

\

\

This results
when we pursue the idea that {\em Kolmogorov complexity measures intrinsic  randomness\/}
and apply it to {\em Bell correlations\/}:
If we have  access to some ``super-Turing machine,'' we let that machine choose
the particular
measurement bases for the two parts of an entangled
quantum system. The  correlations then imply 
 the
sequence of measured values to be 
Turing-uncomputable as well.  This analysis~\cite{nlwc} resembles well-known arguments but 
replaces randomness by complexity.  The new approach has
two conceptual advantages. The first is its
{\em non-counterfactuality\/}:
We do
not  talk about the outcomes of {\em unperformed\/} measurements. (Any Bell inequality does.) The second  is 
{\em context-independence\/}: No process-based notion of randomness is required. Such definitions are typically
embedded into a {\em causal structure\/}~---  but the non-local
correlations {\em themselves}, with  their inexplicability by a
``reasonable'' mechanism within such a structure, are among the
strongest arguments {\em against\/}
 fundamental causality.
The alternative is to consider ``free will'' (randomness, if you will)
as more fundamental
and causality as emerging from it through: ``If
$Y$ is correlated with $X$, and $X$ is {\em freely random}, then~$Y$
must be in $X$'s future.''~\footnote{Already in the early modern
  period there has been a vivid debate about the fundamentality of
  absolute spacetime, for which {\em Newton\/} was  advocating;
{\em Leibniz}, on the other hand, rejected the notion as absurd and
understood spacetime as merely relational, emergent, and of logical
nature. The corresponding debate~\cite{cl} has been decided
by the course of history
 in favor of
Newton~--- with certain exceptions, notably {\em Ernst Mach}.
We are not the only ones to suggest that 
the  choice taken then should be reconsidered today in the light of new insight. 

Although Einstein's
relativity's 
crystallization point was {\em Mach's principle\/}~--- ``inertial
forces are {\em relational}, and not with respect to absolute
spacetime''~---
it does not obey it: For Einstein, spacetime {\em is\/} fundamental
and even in the massless universe, there is the flat spacetime of special relativity.
However, an inherent refutation of rigidly causal thinking {\em is\/} 
contained
in the field equations of general relativity~--- having solutions 
in the
form of {\em closed spacetime curves\/}~\cite{goedel}.  When causality
is dropped, one risks antinomies.
Ruling out the latter does, however, not throw us back to
causality~\cite{nparties}. This observation
has recently been extended to {\em computational
  complexity\/}~\cite{cotton16}:
Non-causal 
circuits  avoiding antinomies are strictly stronger than rigidly causal
circuits.

The opposition between Newton and Leibniz can be seen as the modern
version of the tension between the thinking styles of the pre-Socratic 
philosophers {\em Parmenides\/} and {\em Heraclitus\/}: For the former, all
time and change are pure {\em illusions}, whereas for the latter, {\em
  only 
 change is real\/}: ``You cannot step into the same river twice.''
Nietzsche~\cite{nietzsche} compared the ``cold logician'' Parmenides to {\em ice,}
whereas Heraclitus is the {\em fiery\/} physicist. The postmodern 
manifestation of the tension is the gap between Landauer's
``Information is Physical'' and Wheeler's ``It from Bit.''
In this note, we play on just that opposition  for obtaining our reading of the
second law.}
Then the arrow of time appears as an accumulation of 
  irreversible binary decisions.
 The reverse transformation of such splits
 is 
  not logically reversible and, hence, violates the second law: 
\[{\rm Logical\ Reversibility}\  +\  {\rm Randomness}\  =\  {\rm Thermodynamic\
  Irreversibility\ .}
\]
This equation suggests
how it comes that a law which we  read here  as reversibility is often 
linked to the exact opposite.

\cancel{
xx

\section{Intrinsic Randomness and Spacetime}

TENSION REV/IRREV.
GLEICHUNG: LOG REV + RNESS = PHYS IRREV.
RNESS AND OBJ ARROW OF TIME (other: coarsegr etc.)
EVERETT 
(FEYERABEND)

\section{Philosophical Context}
Philosynth.: sec. hersch, bis mach goedel. zeit baerfuss. 
Qcorell
randomness 
bell
allornothing
Reichenbach Causal 
Baerfuss 

``Wenn es \"uber die Zeitvorstellung kein Einverst\"andnis gibt, dann
stehen alle Begriffe zur Disposition, die eine zeitliche Dimension
haben: Freiheit, Fortschritt, Entwicklung, Wachstum~--- alles
wesentliche Begriffe eine freiheitlichen Demokratie, und alle
pl\"otzlich fragw\"urdig.''
(Krieg und Liebe, Wallstein Verlag, 2018)

(WOHIN? NONCAUS RAND. ODER SCHLUSS?)

Whereas for {\em Parmenides of Elea}, time was a mere illusion~--- ``No
was nor will, all past and future null''~---, {\em Heraclitus\/} saw
space-time as the pre-set stage on which his play of
permanent change  starts and ends.  
The follow-up debate~--- two millennia later and three centuries ago~--- 
between {\em Newton\/} and {\em Leibniz\/}
about as how fundamental space and time, hence, {\em causality}, are to be
seen was decided by the course of science in favor of Newton: In
this 
view, space and time can be imagined as
fundamental and given {\em a priori}. (This applies also to
relativity theory, where space and time  get intertwined and dynamic
but  remain fundamental  instead of  becoming purely
relational in the sense of 
{\em Mach's principle}.) 
Today, we have more reason to question a fundamental causal structure~---
such 
as the difficulty of explaining quantum non-local correlations according 
to Reichenbach's principle. So motivated, we 
care to test
 refraining from assuming space-time as initially given; this has a number of consequences and
implications, some of which we address in this text. 

When causality is dropped, the usual definitions of randomness stop
 making sense. Motivated by this,
we test the use of intrinsic, context-independent ``randomness'' 
measures  such as a string's length minus its (normalized)
fuel value. 
We show that under the Church/Turing hypothesis, Kolmogorov 
complexity relates to this value. We argue that with respect to quantum
non-locality, complexity allows for a reasoning that avoids comparing 
results of different measurements that cannot all be actually carried
out, {\em i.e.}, that is {\em not counterfactual}. Some may see this as a 
conceptual simplification. It also leads to an all-or-nothing flavor 
of the Church/Turing hypothesis: {\em Either no physical system can
generate uncomputable sequences, or even a single photon can}.
Finally, it is asked whether {\em logical
reversibility\/} is connected to the second law of thermodynamics~---
interpreted here in complexities and independent of any context expressed through
probabilities
or ensembles~--- 
and potentially to the arrow of time, past and future.
Finally, we have speculated that if
a causal structure is not
fundamental, how it may emerge from  data-compressibility relations.

When causality is dropped, one risks antinomies.
We show, in the complexity-based view,  that sticking to logical consistency 
does not restore causality but is strictly weaker. This observation
has recently been extended to {\em computational complexity\/}~\cite{cotton16}:
Circuits solely avoiding antinomies are strictly stronger than causal circuits.

---------

Landauer's revised principle puts forward two ideas: First, the erasure
cost is an {\em  intrinsic, context-free, physical measure for 
  randomness\/}
(entirely independent of probabilities and counterfactual statements of
the form ``some value {\em could\/} just as well have been
{\em different\/}'').
The idea that the erasure cost --- or the Kolmogorov complexity
related to it --- is a measure for randomness (independent of
probabilities) can be tested in a context in which randomness
has  been paramount: {\em Bell correlations\/}~\cite{bell} predicted by quantum theory. 
In a proof of principle, it was shown~\cite{nlwc} that in essence, 
a similar mechanism as in the probabilistic setting  arises: {\em If\/}~the
correlation is non-local and the inputs 
are incompressible and non-signaling holds, {\em then\/} the outputs must be
highly complex  as well.
This allows for a discussion of quantum correlations without
the usual counterfactual 
arguments used in derivations of {\em Bell inequalities\/} (combining
in a single formula 
results of different measurements that cannot actually  be 
carried out together). 
 Furthermore, this opens the door to novel
functionalities, namely {\em complexity
  amplification and expansion\/}~\cite{charles}. 
What results is an {\em all-or-nothing flavor of the Church/Turing
  hypothesis\/}: Either no physical computer exists  that is able to produce
non-Turing-computable data ---
or even a ``device'' as simple as a single photon can.

---------

CT 2nd LAW
[fussnote everett; interpret, feyer]
 aus synthese wheeler/landauer 
[fussnote parmenides etc., mach]
CT selbst? Bell!
die sicht mit kompl statt random ist nicht nur luxus sondern not falls
kausal faellt. und die steht in frage wegen bell selbst!
aber auch mach/einstein/goedel. 
2. hs, randomness als arrow.

}

\ \\

\noindent
{\bf Acknowledgments.}
The author thanks 
Mateus Ara\'ujo, Veronika Baumann, \"Amin Baumeler, Charles B\'edard,
Claus Beisbart, Gilles Brassard,
Harvey Brown, Xavier Coiteux-Roy, Caslav Brukner, Harry Buhrman, Matthias Christandl, 
Fabio Costa, Bora Dakic, Fr\'ed\'eric Dupuis, 
Paul Erker, Adrien Feix, J\"urg Fr\"ohlich, Nicolas Gisin, 
Arne Hansen,  Juraj Hromkovi\v{c}, Marcus Huber, 
Alberto Montina, Samuel Ranellucci, Paul Raymond-Robichaud, Louis
Salvail, L.~Benno Salwey,  Andreas Winter, and Chris W\"uthrich
for inspiring discussions. 
{\em --- Grazie mille!}

	Our work is supported by the Swiss National Science
        Foundation (SNF), the National Centre of Competence in
        Research ``Quantum Science and Technology'' (QSIT), and the
{\em Hasler Foundation}.

\end{document}